\documentclass[%
 aip,
 amsmath,amssymb,
 reprint,%
]{revtex4-1}

\usepackage{graphicx}
\usepackage{dcolumn}
\usepackage{bm}

\usepackage[utf8]{inputenc}
\usepackage[T1]{fontenc}
\usepackage{mathptmx}
\usepackage{float}

\begin{document}

\preprint{AIP/123-QED}

\title[Mitigation of Rayleigh-like waves via multi-layer metabarriers]{Mitigation of Rayleigh-like waves in granular media via multi-layer resonant metabarriers\\}

\author{R. Zaccherini}
 \affiliation{ Department of Civil, Environmental and Geomatic Engineering, ETH Z\"urich, Z\"urich 8093, Switzerland}
 \email{zaccherini@ibk.baug.ethz.ch.}
\author{A. Palermo}
 \affiliation{Department of Civil, Chemical, Environmental and Materials Engineering - DICAM, University of Bologna, Bologna 40136, Italy}
\author{A. Marzani}
 \affiliation{Department of Civil, Chemical, Environmental and Materials Engineering - DICAM, University of Bologna, Bologna 40136, Italy}
\author{A. Colombi}
 \affiliation{ Department of Civil, Environmental and Geomatic Engineering, ETH Z\"urich, Z\"urich 8093, Switzerland}
\author{V. Dertimanis}
 \affiliation{ Department of Civil, Environmental and Geomatic Engineering, ETH Z\"urich, Z\"urich 8093, Switzerland}
\author{E. Chatzi}
 \affiliation{ Department of Civil, Environmental and Geomatic Engineering, ETH Z\"urich, Z\"urich 8093, Switzerland}
 \email{zaccherini@ibk.baug.ethz.ch.}

\date{\today}
\begin{abstract}
In this work, we experimentally and numerically investigate the propagation and attenuation of vertically polarized surface waves in an unconsolidated granular medium equipped with small-scale metabarriers of different depth, i.e., arrays composed of one, two, and three embedded layers of sub-wavelenght resonators. Our findings reveal how such multi-layer arrangement strongly affects the attenuation of the surface wave motion within and after the barrier. When the surface waves collide with the barriers, the wavefront is back-scattered and steered downward underneath the oscillators. Due to the stiffness gradient of the granular medium, part of the wavefield is then rerouted to the surface level after overcoming the resonant array. Overall, the in-depth insertion of additional layers of resonators leads to a greater and broader band wave attenuation when compared to the single layer case. 
\end{abstract}

\maketitle

Locally resonant metamaterials can manipulate the propagation of waves at different scales, ranging from the nano-scale, for the control of thermal vibrations \cite{Bruce,Honarvar,Bing,Narayana}, up to the geophysical scale, for the attenuation of seismic vibrations \cite{Finocchio,Casablanca,Colombi,Colquit,Bursi,Achaoui,Roux,Maurel}. They consist of resonant building blocks of sizes that are significantly smaller than the wavelengths of the phenomena they affect \cite{Hussein}. Arrays of meter-scale mechanical oscillators, referred to as metabarriers \cite{Vasilis,Antonio1,AndreaBarrier}, have recently been proposed for the mitigation of surface ground motions. When embedded in homogeneous substrates, these resonant barriers can open bandgaps in the surface wave frequency spectrum \cite{Antonio1}. The bandgaps arise from the coupling between the incoming surface waves and the resonances of the metabarrier, which leads to a surface-to-shear wave conversion. As a result of this energy delocalization, the surface motion can be significantly attenuated. The frequency spectrum and spatial extent of surface-to-shear wave conversion is highly influenced by the stratigraphy of the soil \cite{Pu}. In particular, when the resonant array is embedded in a medium with an inhomogeneus depth-dependent stiffness profile, the wave conversion mechanism may be hampered. Experimental evidences of such phenomenon have been observed in metasurfaces consisting of a single layer of sub-wavelength oscillators, and interacting with guided surface acoustic modes (GSAMs) propagating at the surface of an unconsolidated granular medium \cite{Palermo,Zaccherini}. Such GSAMs originate due to the peculiar gravity-induced shear $G$ and the bulk $B$ elastic moduli profiles of the granular material \cite{Gusev}, which exhibit a power-law dependency on the compacting pressure $p= \rho gz$, i.e., $G, B \propto p^{\alpha}$, where $\rho$ is the medium density, $g$ the gravitational constant, and $z$ the depth. These surface acoustic modes consist of vertically polarized (P-SV) waves, composed of continuously interacting longitudinal (P) and shear vertical (SV) modes, and shear horizontal (SH) waves \cite{Aleshiv}. The studies confirmed that the classical surface-to-shear conversion phenomenon, observed in homogeneous media, is inhibited by the gravity-induced stiffness profile of the material.

In this work, we experimentally and numerically investigate the propagation of P-SV surface modes in a granular medium equipped with three small-scale metabarriers, consisting of arrays of vertical sub-wavelength oscillators arranged across multiple layers along the medium. Our aim is to investigate whether such a multi-layer disposition can enhance the overall wave attenuation. 
\begin{figure*}
\includegraphics[width=1\textwidth]{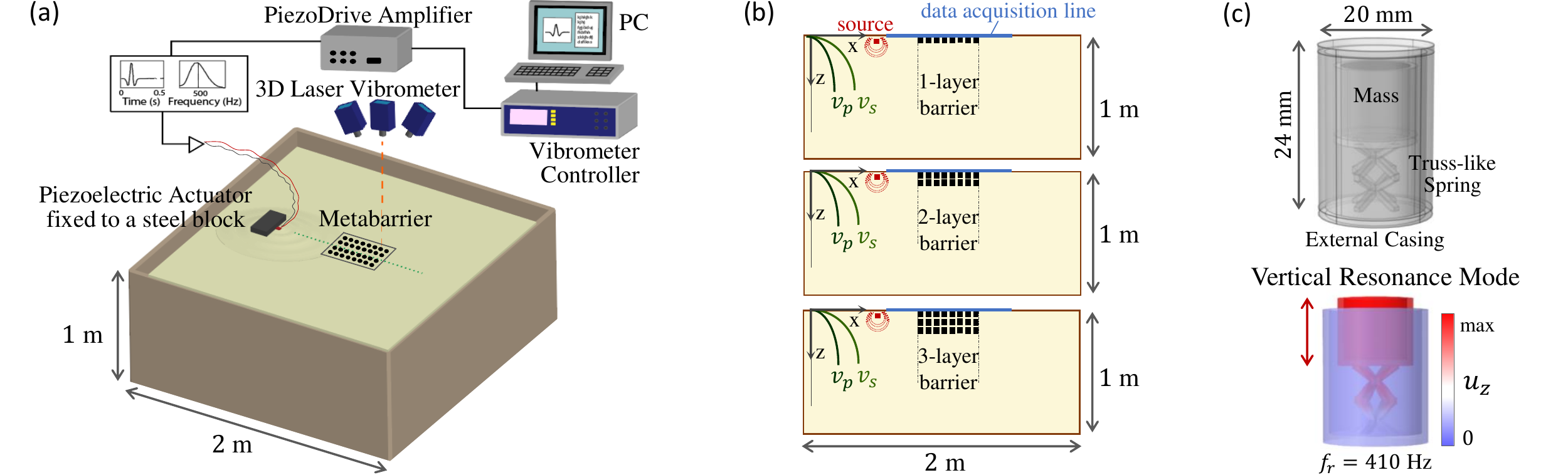}
\caption{\label{fig:Fig1} (a) Schematic of the experimental setup. (b) Schematic of the vertical cross-section of the box for the three investigated metabarrier configurations. (c) Model of the vertical oscillator together with its first resonant shape mode.}
\end{figure*}
To this end, we built the experimental setup shown in Fig.~\ref{fig:Fig1}(a). It comprises a wooden box (2000${\times}$1500${\times}$1000 mm) filled with granular material (150-${\mu}$m-diameter glass beads with density of 1600 Kg/m$^3$), a Polytech PSV-500 3D laser Doppler vibrometer, and a three-axis piezoelectric actuator driven by a PiezoDrive amplifier. Similar setups have been used to study the propagation of surface acoustic modes in granular media \cite{Bodet,Jacob,Bodet2} and to investigate their interaction with arrays of vertical \cite{Palermo} and horizontal \cite{Zaccherini} oscillators. One at a time, we embed below the surface of the granular medium, three barriers constituted by one, two and three layers, respectively, of sub-wavelength vertical oscillators arranged according to a 4${\times}$8 rectangular grid (see Fig.~\ref{fig:Fig1}(b)). Each resonant unit, depicted in Fig.~\ref{fig:Fig1}(c), comprises a rigid cylindrical shell and a truss-like spring, both 3D-printed in Acrylonitrile Butadiene Styrene (ABS), and a 12 g cylindrical mass made out of steel. This set of resonators was originally realized in \cite{Palermo} to analyse the interaction between the P-SV waves and a single layer metabarrier. The resonant frequency is set, by design, at $410$ Hz, when the oscillators are embedded below the granular surface. The piezoelectric actuator generates P-SV surface modes. As input signal, we employ a Ricker wavelet centered at $500$ Hz to realize a pulse-like excitation. For all tested configurations, namely pristine granular medium and in-depth metabarriers, we use the 3D vibrometer to record the particle vertical velocity along the symmetry axis of the box with a constant step of $\Delta x=8$ mm (see the dotted green line of Fig.~\ref{fig:Fig1}(a)).
\begin{figure*}
\includegraphics[width=1\textwidth]{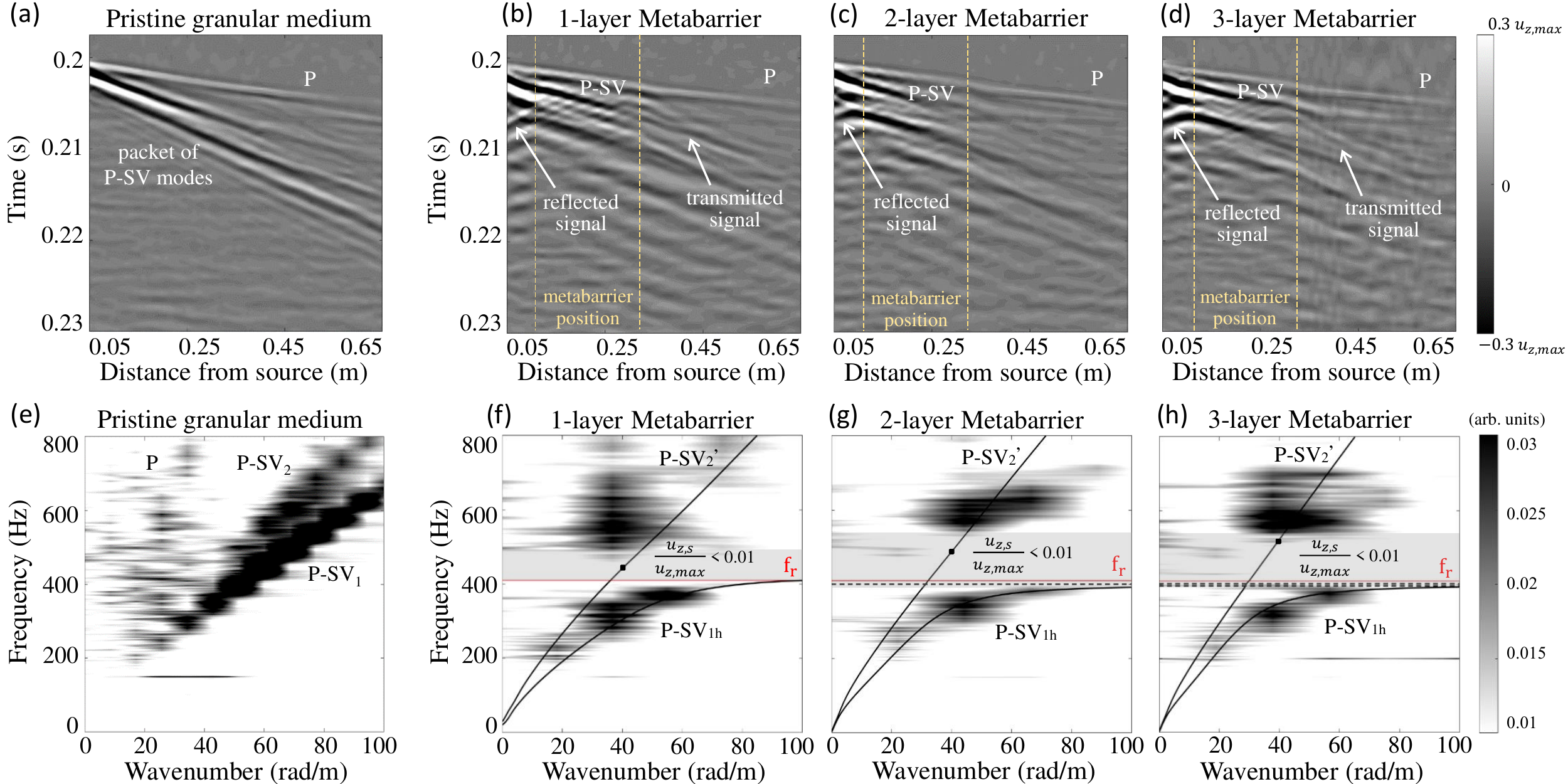}
\caption{\label{fig:Fig2} Top: Seismograms of a Ricker pulse propagating in the pristine granular medium (a) and through a 1-layer (b), 2-layer (c), and 3-layer (d) metabarrier, respectively. Bottom: Experimental (dark spots) and numerical (black line) dispersion curves of P-SV modes propagating in the pristine granular medium (e) and interacting with a 1-layer (f), 2-layer (g), and 3-layer (h) metabarrier, respectively. The red lines represent the oscillator resonant frequency, while the dashed black lines correspond to localized resonance modes within the mechanical oscillators.}
\end{figure*}

Figure \ref{fig:Fig2}(a) shows the seismogram generated by a Ricker pulse propagating in the pristine granular medium, obtained from the data collected along the blue line illustrated in Fig. \ref{fig:Fig1}(b). Two distinct wavetrains mainly dominate the seismogram: a quasi-compressional (P) wave, discarded in our study, and a packet of P-SV surface waves that propagates with lower velocity. After application of the two-dimensional (2D) discrete Fourier transform (DFT) to the time waveforms, we obtain the frequency-wavenumber spectrum (see Fig. \ref{fig:Fig2}(e)). The latter highlights the frequency content of the acoustic modes dominating the seismograms: 400-800 Hz for the P wave and 200-800 Hz for the two slowest P-SV waves (labelled as "P-SV$_1$" and "P-SV$_2$"). Figure \ref{fig:Fig2}(b) shows the seismogram obtained from the data collected along the blue line, when a 1-layer metabarrier is embedded below the medium surface. The two dashed yellow vertical lines indicate the first and the last resonators lines, respectively. When the surface modes impinge on the resonant barrier, new slower direct and reflected wavetrains arise. The frequency-wavenumber spectrum, displayed in Fig.~\ref{fig:Fig2}(f), shows that the low frequency components of the signal, associated with the lowest-order mode, strongly couple with the collective resonances of the metabarrier, generating a flat hybrid mode (here labelled as “P-SV$_{1h}$”), which converges asymptotically to the oscillator resonance frequency (f$_r$=410 Hz, red line). As a result of this hybridization mechanism, a region from 410 Hz to around 500 Hz, where the wave amplitude significantly drops, can be distinguished. The higher-frequency components of the signal, traveling at a higher velocity, are associated with the second-order mode P-SV$_{2}$', which appears scattered by the resonator casings. Similar results are found when analysing the interaction between the P-SV modes and the 2- and 3-layer metabarriers (see the related seismograms of Figs.~\ref{fig:Fig2}(c)-(d) and the dispersion relations of Figs.~\ref{fig:Fig2}(g)-(h)). In these configurations, we observe that the amplitude of the wave reflection in front of the metabarrier intensifies by increasing the number of resonator layers. Similarly, the wave attenuation bandwidth significantly enlarges when a second in-depth layer of resonators is added (2-layer metabarrier). This result can be ascribable to the coupling between the two layers of oscillators, which slightly alters the frequency associated with the local resonances \cite{Lemoult}. No significant differences are found in the frequency-wavenumber spectrum when introducing a third layer of resonators.
\begin{figure}
\includegraphics[width=0.5\textwidth]{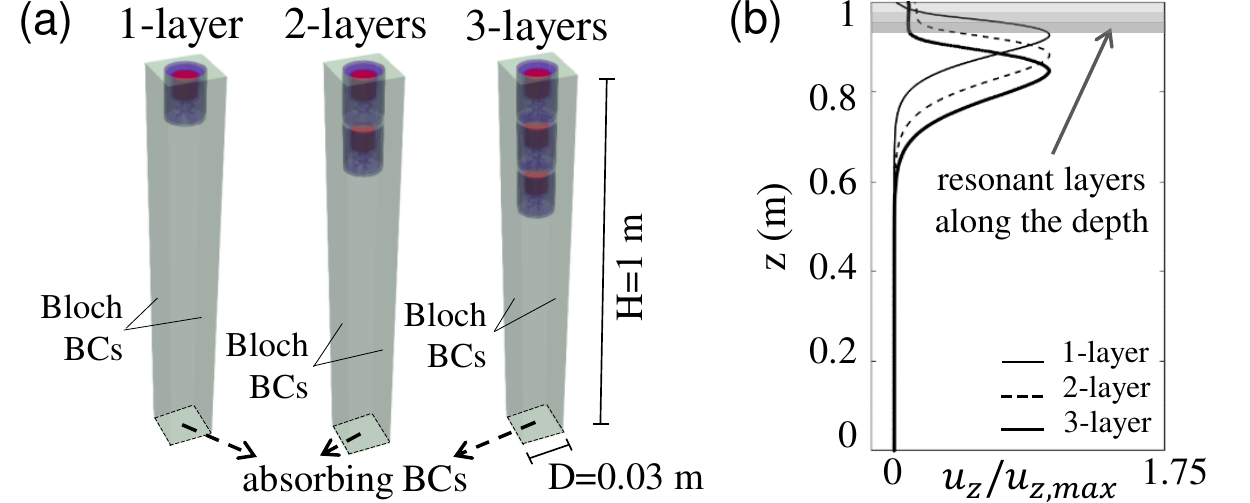}
\caption{\label{fig:model} (a) Schematic of the 3D unit cells with 1, 2, and 3 in-depth layers of resonators developed in COMSOL Multiphysics®. (b) Vertical displacement component of the second-order surface mode interacting with the 1, 2 and 3-layer metabarriers for $k = 41.1$ rad/m.}
\end{figure}
To gain a better insight into the dynamics of P-SV waves interacting with the three metabarriers, we numerically calculate the dispersion curves. To this purpose, we model the three-dimensional (3D) unit cells of the three metabarriers with the finite-element (FE) software COMSOL Multiphysics® (see Fig.~\ref{fig:model}(a) and Supplementary Material for further details). We impose periodic Bloch boundary conditions and perform eigenfrequency analyses by sweeping the wavenumber in the range 0 - $k_{max}$ (with $k_{max}$ = $\pi/D$ = 105 rad/m, where $D$ represents the unit-cell width). The numerical dispersion curves, depicted as solid black lines in Figs.~\ref{fig:Fig2}(f)-(g)-(h), well match the experimental energy distribution of the P-SV waves interacting with the mechanical oscillators. Flat curves (depicted as black dashed lines in Figs.~\ref{fig:Fig2}(g)-(h)), which correspond to localized resonance modes within the mechanical oscillators, appear in the dispersion relations of the 2 and 3-layer metabarriers. The frequencies of these localized modes slightly differ from each other, confirming the coupling between the stacked oscillators \cite{Lemoult}. The numerical findings reveal that the frequency gap vanishes, as the P-SV$_2$' mode does not hybridize with the local resonant modes.
Only the P-SV$_1$ mode, which exhibits a significant vertical displacement component at the surface level (see Fig. S1 in the Supplementary Material), strongly couples to the resonators and hybridizes. However, an insight on the displacement field of the P-SV$_2$' mode (Fig.~\ref{fig:model}(b)) reveals an extended frequency range in proximity to the natural frequency of the resonators, where the surface displacement nearly vanishes, i.e., u$_{z,s}$/u$_{z,max}$ < 0.01, with u$_{z,s}$ being the surface displacement and u$_{z,max}$ the maximum displacement along depth. The shaded gray areas of Figs.~\ref{fig:Fig2}(f)-(g)-(h) identify these frequency regions, which well match the energy gaps detected experimentally. Note that the insertion of further in-depth layers of resonators shifts the location of the maximum displacement at greater depths, thus extending the region with attenuated displacement field (see Fig.~\ref{fig:model}(b)).
\begin{figure}
\includegraphics[width=0.5\textwidth]{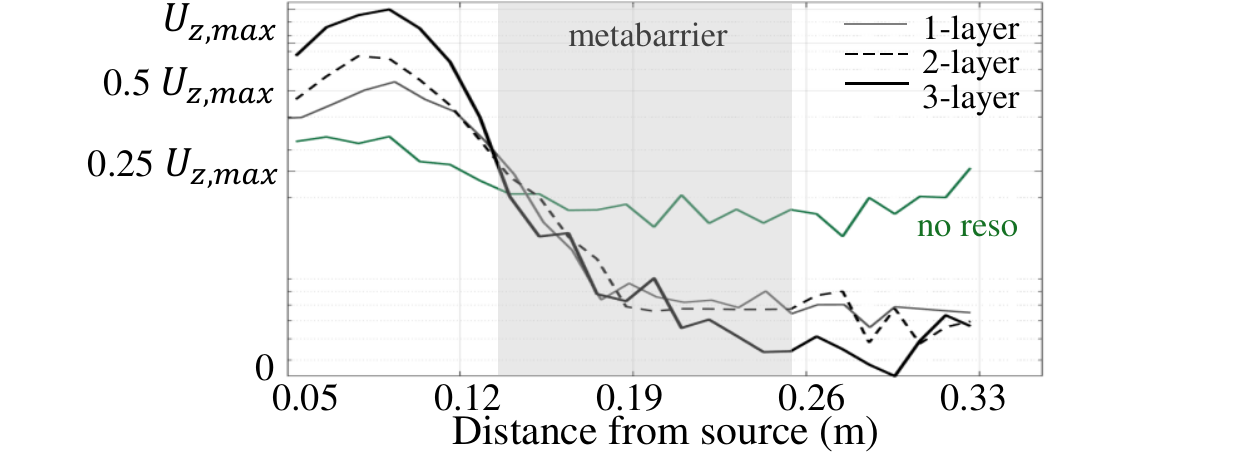}
\caption{\label{fig:Fig3} Experimental spectral surface displacement (logarithmic scale) computed in the granular medium without and with the three metabarriers.}
\end{figure}
\begin{figure*}
\includegraphics[width=1\textwidth]{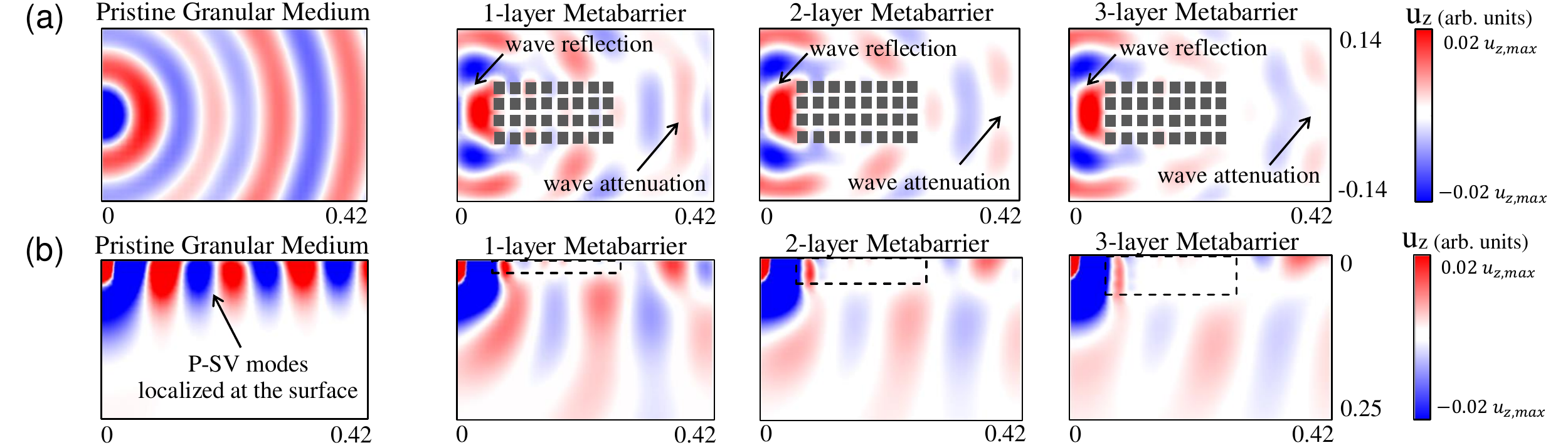}
\caption{\label{fig:Fig5} (a) Equal time snapshots of the displacement field computed numerically of P-SV waves propagating in the pristine granular medium and interacting with the three in-depth metabarriers at t$_1$ = 42.0 ms. (b) Vertical cross sections through the center of the model, showing the numerical displacement field of P-SV waves traveling in the pristine granular medium and through the in-depth metabarriers at t$_2$ = 42.7 ms.}
\end{figure*}

We now compare the amplitude of the waves propagating at the surface of the pristine granular medium with the amplitude of the waves interacting with the three metabarriers. To this purpose, we obtain displacement time histories from the velocity data recorded at the surface for an array of 25 points equispaced along the blue line illustrated in Fig.~\ref{fig:Fig1}(b). The numerical integration is performed using the trapezoidal rule. For each acquisition point along the surface line, we then calculate the maximum of the spectral displacement inside the frequency range 430-500 Hz. The results are shown in Fig.~\ref{fig:Fig3}. The shaded grey area identifies the position of the metabarriers. We observe that the spectral displacement amplitude in front of the metabarriers significantly exceeds the one recorded in the pristine granular medium. Moreover, the amplitude increases as additional in-depth layers of resonators are introduced. Since the intensity of the excitation signal is kept constant for all configurations, this indicates that part of the surface wave is back-reflected by the resonant arrays and this phenomenon intensifies as the depth of the metabarrier increases. By comparing the spectral amplitude within and behind the resonant arrays, we note that the amplitude in the pristine granular medium approximately doubles the one measured with the metabarriers. Additionally, we observe that the amplitude reduces as the depth of the barrier increases to become almost null when a third layer of oscillators is introduced. 

To gain further insight into the in-depth propagation of these surface modes interacting with multi-layer barriers, we perform 3D time-transient numerical simulations using SPECFEM3D \cite{spec}, a spectral element based software. In particular, we consider a granular 1-m-deep half-space, 2${\times}$1.5 m wide, modeled as a linear elastic continuum according to the GSAM theory described in Ref. \cite{Aleshiv}. We develop time-domain simulations for each of the experimentally analysed configurations (for further details see the Supplementary Material). Figure~\ref{fig:Fig5}(a) shows the vertical displacement field of P-SV waves propagating in the pristine granular medium and interacting with the three metabarriers at the time instant t$_1$ = 42.0 ms, filtered inside the frequency range of the attenuation zone. In the pristine granular medium, the P-SV wavefronts propagate undisturbed at the surface. Conversely, as the P-SV modes impinge the resonant arrays, the wavefronts are splitted. As observed in the experiments, the wave attenuation after the metabarriers intensifies as the number of in-depth resonant layers increases. Moreover, a reverse-concavity wavefront prior to each metabarrier, hallmark of wave reflection, is visible. The vertical cross sections (Fig.~\ref{fig:Fig5}(b)), cut through the middle of the models, illustrate the P-SV wavefield. The black dashed squares mark the position of the metabarriers. We observe that the portion of signal energy, which is not reflected backward, separates from the surface and is channeled underneath the resonators. The higher-order P-SV modes, which do not hybridize with the oscillators, travel therefore below the resonators, without being converted into leaky shear waves, to then return to the surface after overcoming the metabarrier. Although the surface-to-shear conversion mechanism is prevented by the continuous increase of the elastic profile of the granular medium, the P-SV waves, which return to the surface, are significantly attenuated.
\begin{figure}
\includegraphics[width=0.5\textwidth]{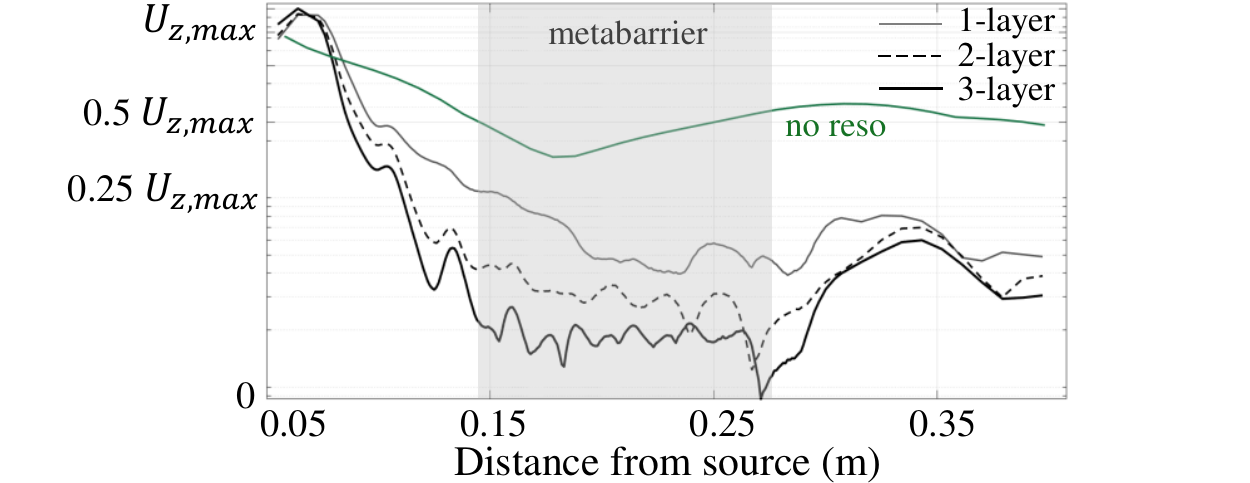}
\caption{\label{fig:Fig8} Numerical spectral surface displacement (logarithmic scale) computed in the granular medium with and without the three metabarriers.}
\end{figure}
To quantify this attenuation, we calculate the maximum of the spectral displacement at the surface along the symmetry line of the model, inside the frequency range 430-500 Hz, with and without the metabarriers (see Fig.~\ref{fig:Fig8}). As observed in the experimental results shown in Fig.~\ref{fig:Fig3}, when inserting the oscillators below the granular surface, the wave amplitude increases in front of the mebatarrier and decreases within and behind it. By increasing the depth of the metabarrier, both wave reflection and attenuation rise. However, the displacement amplitudes obtained numerically in front of the multi-layer barriers are less pronounced than those recorded in the experiment. These differences, which can be found by quantitatively comparing the experimental and numerical results, may be partially ascribed to the the damping of the granular medium, which is neglected in the numerical model. Behind the metabarriers, the amplitude slightly increases, which confirms that the stiffness gradient of the granular medium preserves the surface energy confinement. However, we also note that the amplitude in the pristine medium is approximately twice that measured with the 3-layer barrier, which demonstrates that the addition of in-depth resonators strongly enhances the overall surface wave attenuation.

To conclude, we experimentally and numerically analyse the wave amplitude attenuation induced by three different metabarriers, composed of one, two, and three in-depth layers of sub-wavelength oscillators, respectively, embedded in an unconsolidated granular medium with a depth-increasing stiffness profile. The findings reveal that the addition of in-depth resonators strongly affects the wave amplitude at the surface. In particular, dispersion and mode shape analysis reveal an attenuation frequency range, whose bandwidth considerably widens by inserting a second in-depth layer of resonators, where the surface wave amplitude within the metabarrier region nearly vanishes. Experiments and time-domain simulations confirms that when the surface waves collide with the resonant array, part of the energy is back-scattered and part is conveyed downward underneath the oscillators. Behind the metabarrier, the wave returns to the surface level, but its amplitude is considerably decreased with increasing depth of the metabarrier. Since we expect that similar outcomes may be observed in stratified soils that exhibit a power-law elastic profile at the geophysical scale, we believe that these findings can serve as a base for the design of resonant isolation barriers in heterogeneous substrates. 

\newpage

\section*{Supplementary Material}
\label{section:Supplementary Material}
See the supplementary material for the detailed descriptions of the experimental setup preparation and experimental procedures, the propagation of the surface modes in the pristine granular medium, the numerical dispersion analysis developed in COMSOL Multiphysics®, and the time-domain numerical simulation performed in SPECFEM3D.

\begin{acknowledgments}
This research was partially supported by the ETH Research Grant No. 49 17-1 to E.C., the Ambizione fellowship PZ00P2-174009 and the H2020 FET Open project BOHEME (grant agreement No. 863179) to A.C.
\end{acknowledgments}

\section*{Data Availability}
The data that support the findings of this study are available from the corresponding author upon reasonable request.\\
\\

\bibliographystyle{unsrt}

\nocite{*}
\bibliography{aipsamp}

\end{document}